\documentclass[%
 reprint,%
 amssymb, amsmath,%
aip,
jap,%
 superscriptaddress,%
]{revtex4-1}

\usepackage{graphicx}
\usepackage{stmaryrd}
\usepackage{color}
\usepackage{bm}%
\usepackage[colorlinks=true,linkcolor=blue]{hyperref}%
\expandafter\ifx\csname package@font\endcsname\relax\else
 \expandafter\expandafter
 \expandafter\usepackage
 \expandafter\expandafter
 \expandafter{\csname package@font\endcsname}%
\fi
\usepackage{longtable}
\usepackage[utf8]{inputenc}
\usepackage{epstopdf}

\begin{document}

\title{Generalized interface models for transport phenomena: unusual scale effects in composite nanomaterials}%

\author{Fabio Pavanello}
\affiliation{IEMN, UMR CNRS 8520, Avenue Poincar\'{e}, BP 60069, 59652 Villeneuve d'Ascq, France}
\author{Fabio Manca}
\affiliation{Dipartimento di Fisica, Università di Cagliari, Cittadella Universitaria, 09042 Monserrato, Cagliari, Italy}
\author{Pier Luca Palla}
\affiliation{IEMN, UMR CNRS 8520, Avenue Poincar\'{e}, BP 60069, 59652 Villeneuve d'Ascq, France}
\author{Stefano Giordano}
\email{Stefano.Giordano@iemn.univ-lille1.fr}
\affiliation{International Associated Laboratory LEMAC:\\
IEMN, UMR CNRS 8520, PRES Lille Nord de France, ECLille, Avenue Poincar\'{e}, BP 60069, 59652 Villeneuve d'Ascq, France}

\begin{abstract}
The effective transport properties of heterogeneous nanoscale materials and structures  are affected by several geometrical and physical factors. Among them the presence of imperfect interfaces plays a central role being often at the origin of the scale effects. 
To describe real contacts between different phases some classical schemes have been introduced in literature, namely the low and the high conducting interface  models. Here, we introduce  a generalized formalism, which is able to take into account the properties of both previous schemes and, at the same time, it implements more complex behaviors, already observed in recent investigations. 
We apply our models to the calculation of the effective conductivity in  a paradigmatic structure composed of a dispersion of particles.
In particular we describe the conductivity dependence upon the size of the inclusions finding an unusual non-monotone scale effect with a pronounced peak at a given particle size. We introduce some intrinsic length scales governing the universal scaling laws. 
\end{abstract}

\maketitle

\section{Introduction}
One of the central problems in material science is to evaluate the effective electric, magnetic, elastic and thermal properties governing the  physical behavior of heterogeneous materials.\cite{torquato,milton}
In recent years, with the  progressive miniaturization of structures and devices, possible size effects have attracted an ever increasing interest. One crucial property that usually drives the scale effects in structured materials is the complexity of interfaces between different phases. Typically, in the macroscopic modeling, the interfaces are assumed to be perfect. In the context of the electrical conduction it means that  the
potential $V$ and the normal component of the current density $\vec{J}$ are continuous across any interface:\cite{stratton,landau}  $ \llbracket V \rrbracket = 0 $ and $ \llbracket \vec{J}\cdot\vec{n} \rrbracket = 0 $, where the symbol $ \llbracket f \rrbracket$ represents the jump of the function $f$ across the interface. 
This approximation turns out to be valid in the case of small surface/volume ratio.
However, in many real cases of technological interest, e.g. nanocomposites, it is important to take into consideration the specific properties of the contacts among the constituents. To this aim, two effective interface models have been so far introduced for describing two extreme situations in a zero thickness formulation. Moreover, other models are based of an explicit interphase of finite thickness and, therefore, they typically consider a three-phase heterogeneous material composed of the inclusions, the interphase medium and the matrix.\cite{benv-jmps,benv-proc,gu-he}

\begin{figure*}[t]
\includegraphics[width=7cm]{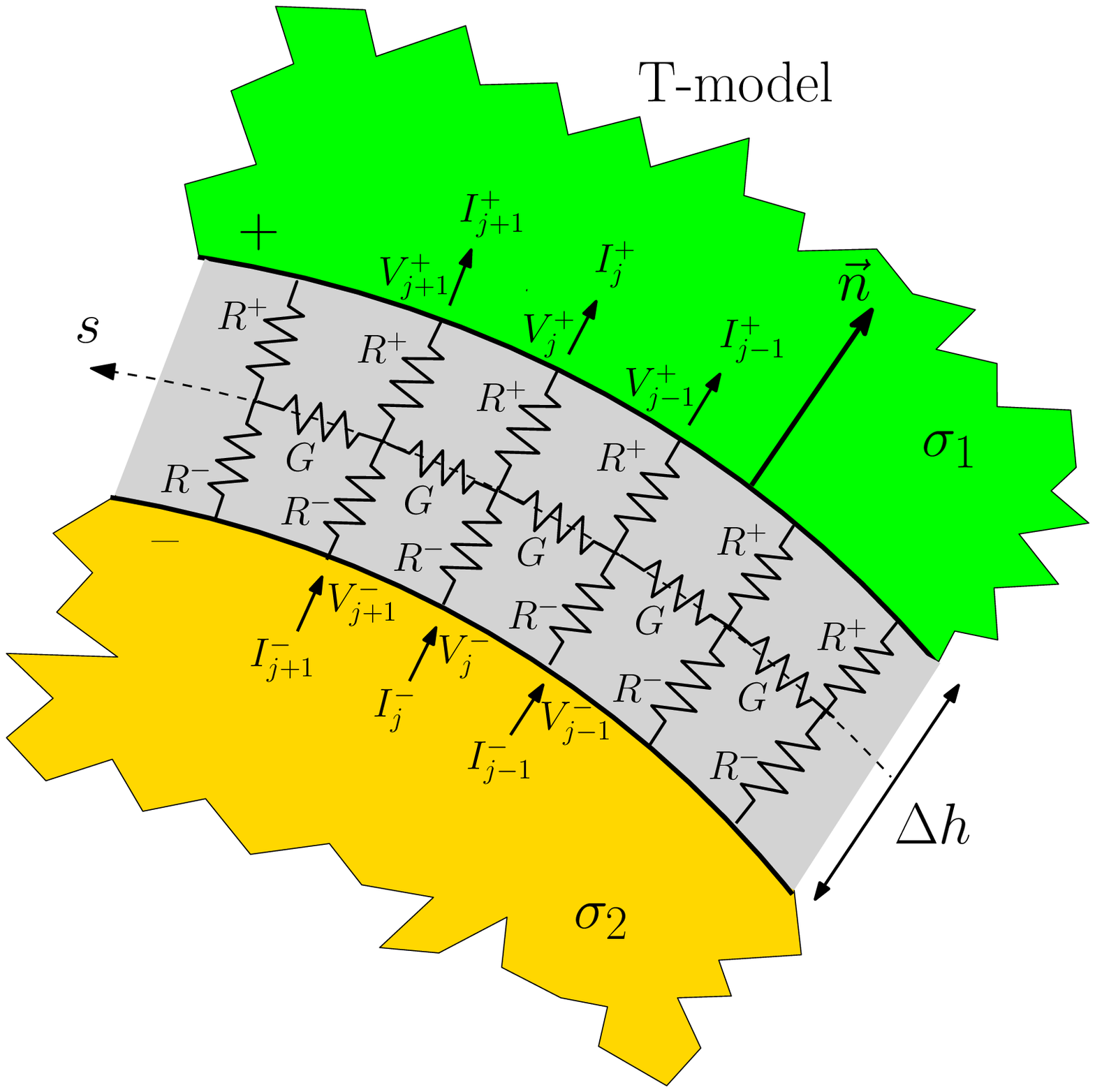}\,\,\,\,\,\,
\includegraphics[width=7cm]{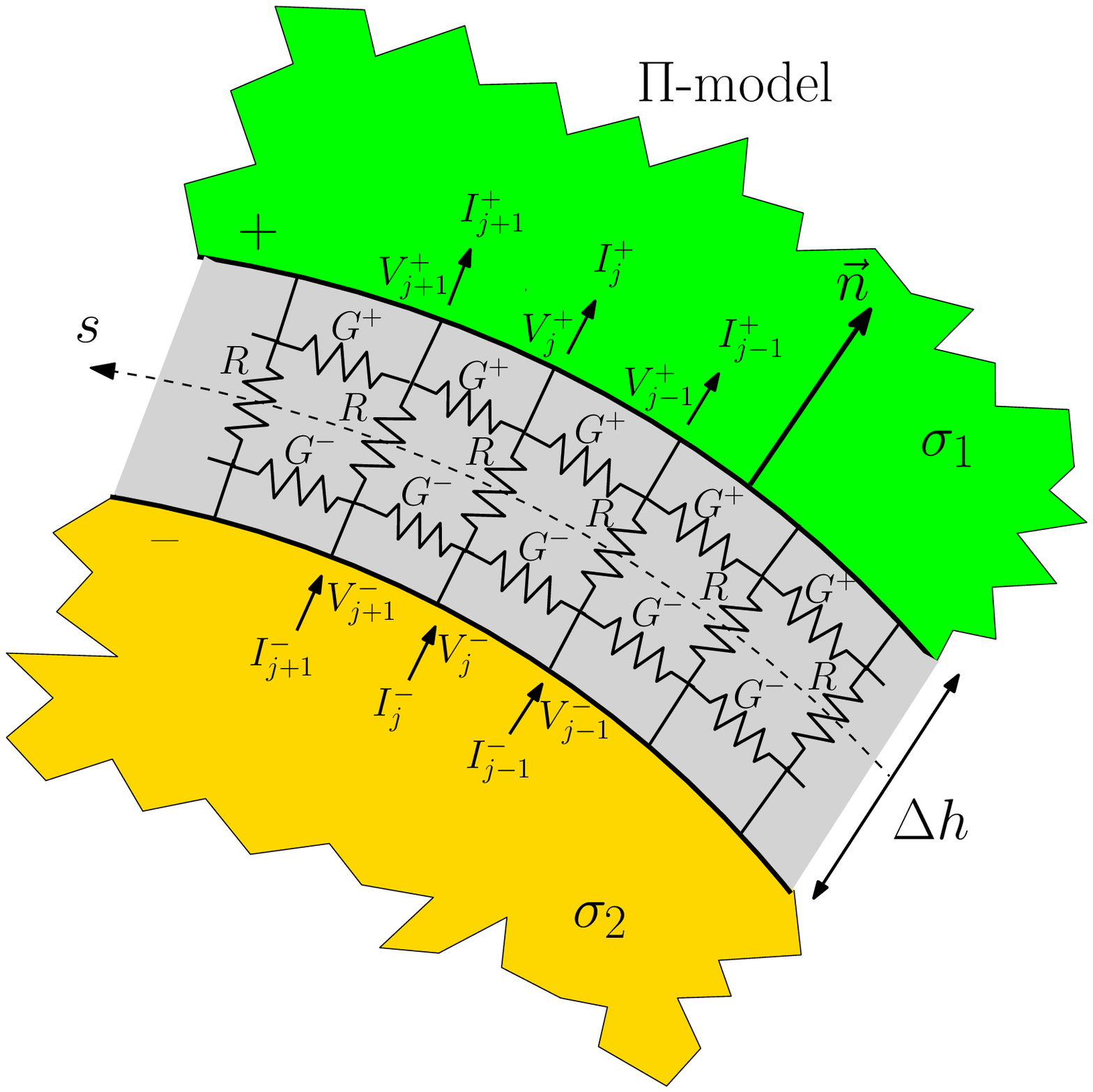}
\caption{\label{figure1}(Color online) Schemes of the dual anisotropic imperfect interfaces (T-model and $\Pi$-model) between two homogeneous media with conductivities $\sigma_{1}$ and $\sigma_{2}$. }
\end{figure*}

The first zero thickness model is called \textit{low conducting interface} and it is based on the Kapitza resistance, introduced in the context of the thermal conduction.\cite{Kapitza} 
According to this approach $ \llbracket \vec{J}\cdot\vec{n} \rrbracket = 0 $, while the potential suffers a jump proportional to the local flux, $ \llbracket V \rrbracket = -r \vec{J}\cdot\vec{n}$, where $r$ is the Kapitza-like resistance. 
The second model, called \textit{high conducting interface}, concerns the case of an interphase of very high conductivity with  vanishing thickness. In this situation  
$ \llbracket V \rrbracket = 0 $, while the normal component of the current density is proportional to the surface Laplacian of the potential, $ \llbracket \vec{J}\cdot\vec{n} \rrbracket = g\nabla_{S}^2 V$, where $g$ represents the interphase conductance.
Several investigations on heterogeneous materials with low  \cite{benvmilo,benv,hasse,tor-rin,lip1,lip2,chen,nan,hashin,duan-thermal,bonnet} or   high  \cite{tor-rin,duan-thermal,lip3,lip4,moloh,yvonnet,bonnet-he} conducting interfaces can be found in literature.

In many cases, the behavior of complex interfaces cannot be simply described through the low or high conducting model. In fact, these schemes account for a single interlayer with an extreme (high or low) value of the conductivity, while real interface typically exhibit a complex or multilayered structure. To overcome this difficulty we introduce a generalized anisotropic interface formalism, which consider both the normal resistance (similarly to the Kapitza case) and the tangential conductance (as in the high conducting interface model).
The term anisotropic refers to the fact that the normal resistance and the tangential conductance are completely independent, describing a different behavior in the two directions. As discussed below, in order to integrate both the normal and tangential features, two dual schemes are possible, as shown in Fig.\ref{figure1}. They exploit the classical T and $\Pi$ electric lattice structures. 
By means of this approach we take into account all situations comprised between the low and high conducting interface models, which can be seen as limiting cases of the present theory. 
In our schemes both the potential and the normal component of the current density are discontinuous at the interface. The richness of the proposed models allows us to effectively describe the behavior of real imperfect/multilayered/structured interfaces, which can be found in several heterogeneous materials of technological interest and, in particular, for
those displaying a complex nanoscale structure.

At first, we have applied the generalized interfaces to model a single particle embedded in a different matrix (inhomogeneity).
One of the most important technique used to study  this system is based on the Eshelby formalism.
It has been introduced in the context of the isotropic elasticity theory,\cite{eshelby1,eshelby2} generalized to the anisotropic elasticity\cite{walpole} and applied to the electric, magnetic or thermal case.\cite{taya,lak,giordano} 
A relevant universal property states that the field induced in cylindrical or spherical  particles with zero-thickness low or high conducting interfaces is uniform if the externally applied field is so.
It is true for both isotropic constituents and anisotropic ones, as recently proved.\cite{bonnet,bonnet-he} 
In the present work, we show that the uniformity property of the internal field for spheres and cylinders is preserved also for our generalized interface models.
Moreover, to extend the validity of  the Eshelby approach, we propose a method which is able to determine the field within and around a single particle even if the externally applied field is not uniform.

The previous results have been applied to determine the overall conductivity (through an effective medium theory\cite{giorda,giorda1}) of a dispersion of particles with imperfect interfaces.
We have verified that, in contrast to perfectly bonded inclusions, the effective properties depend upon the size of the inhomogeneities. Interestingly enough, while in the case of low or high conducting interfaces the scale effects are described by  monotone scaling laws, the present generalized models show  non-monotone scale effects with a sizeable peak of the transport properties. This point can be considered as a specific signature of the complex resistive/conductive behavior of the interface. A description of this intriguing behavior has been made through different intrinsic length scales governing the universal scaling laws. Typical material science problems where these interface models can be profitably applied are the following: tailoring of composites with semiconductor whiskers;\cite{appl1} thermal optimization of metal/dielectric interfaces\cite{appl2} and change materials through nanoclusters of stable oxides;\cite{appl3} analysis of thermal and electric conductivity of carbon-based nanostructures;\cite{appl4,appl5} size effects understanding in SiC/epoxy (or similar) nanocomposites.\cite{appl6,appl7}   

Throughout all the paper we develop the formalism with the terminology of the electrical transport, but all results can be applied to the analogous situations of thermal conduction, antiplane elasticity, magnetic permeability and electric permittivity as well.

\section{The dual interface models}

To begin, we introduce a simple lattice network taking into account the normal resistors $R^+$ and $R^-$ and the tangential conductance $G$ (see Fig.\ref{figure1}, T-model). 
This structure is able to consider both the anisotropy (along the normal and tangential directions) and the different behavior of the normal conductivity on the two sides of the interface. 
For the moment we consider a curvilinear interface between two different materials of a planar structure (2D geometry). The generalization to the arbitrary three-dimensional case will be made straightforwardly. 
By a direct application of the Kirchhoff circuit laws we obtain two equalities describing the voltage jump and current jump across the interface (the definition of the relevant quantities is shown in Fig.\ref{figure1})
\begin{eqnarray}
\label{uno}
V_{j}^+ -V_{j}^- &=& -R^+ I_j^+-R^-I_j^-,
\\
\nonumber
I_{j}^+ -I_{j}^- &=& GR^+\left(I_{j-1}^+ -2I_{j}^+ +I_{j+1}^+\right) \\&&+G\left(V_{j-1}^+ -2V_{j}^+ +V_{j+1}^+ \right) .
\label{due}
\end{eqnarray}
In the limit of a continuous zero-thickness interface we easily obtain from Eqs.(\ref{uno}) and (\ref{due}) the relations for the interface in the form 
\begin{eqnarray}
\label{saltoV1}
\llbracket V \rrbracket &=& -r^+ \left( \vec{J}\cdot\vec{n}\right)^+ -r^-\left( \vec{J}\cdot\vec{n}\right)^-,
\\
\label{saltoJ1}
\llbracket \vec{J}\cdot\vec{n} \rrbracket &=& gr^+\frac{\partial^{2}}{\partial s^{2}}\left( \vec{J}\cdot\vec{n}\right)^+ +g\frac{\partial^{2}}{\partial s^{2}}V^{+}.
\end{eqnarray}
The parameters $ r^-, r^+ $ and $g$ are the suitably rescaled counterparts of $R^-$, $R^+$ and $G$ ($ r^-$ and $ r^+ $ are measured in $\Omega$m$^2$ and $g$ in $\Omega^{-1}$).\cite{tor-rin}
In previous expressions, the partial derivatives are performed with respect to the variable $s$, which represents the curvilinear abscissa along the arbitrarily curved interface on the plane. As usual, in the three-dimensional case the operator $\partial^2/\partial s^2$ must be substituted with the surface Laplacian $\nabla_{S}^2$, which is introduced and discussed in Appendix A. We can observe that the present approach reproduces the \textit{low conducting interface} model if $g=0$ (with a Kapitza resistance $r=r^- + r^+$) and the \textit{high conducting interface} model if $r^-=r^+=0$. 

The low conducting model is characterized by a sequence of normal resistances $R=R^+ + R^-$ (T-model with $G=0$). If we consider $\Delta s$ as the step along the curvilinear abscissa $s$ and $\Delta z$ as the step along the direction perpendicular to the plane represented in Fig.\ref{figure1} we have $R=\frac{1}{\sigma_{\perp}}\frac{\Delta h}{\Delta s \Delta z}$ where $ \sigma_{\perp} $ is the normal conductivity of the interphase of thickness $ \Delta h $. The Kapitza resistance is therefore given by $r=R \Delta S$ where $\Delta S=\Delta s \Delta z$ is the area element associated to $\vec{J}\cdot\vec{n}$; we finally obtain $r=\lim_{\Delta h \rightarrow 0, \sigma_{\perp} \rightarrow 0}\frac{\Delta h}{\sigma_{\perp}}$.\cite{tor-rin} Similarly, the high conducting model is characterized by a series of tangential conductances $G$ (T-model with $R^-=R^+=0$). It is simple to observe that $G=\sigma_{\parallel}\frac{\Delta h \Delta z}{ \Delta s}$ where $\sigma_{\parallel}$ is the tangential conductivity of the interphase. The specific conductivity is therefore given by $g=G\frac{\Delta s^2}{\Delta S}$ (where $\Delta S=\Delta s \Delta z$) and we obtain the result $g=\lim_{\Delta h \rightarrow 0, \sigma_{\parallel} \rightarrow \infty}  \sigma_{\parallel}\Delta h$.\cite{tor-rin} So, we have a direct link between the interphase properties ($ \Delta h $, $ \sigma_{\perp} $, $\sigma_{\parallel}$) and the models parameters ($r$, $g$) for the high and low conducting interfaces. Interestingly enough, we observe that when we consider an anisotropic single layer interphase (which is uniaxial or transversely isotropic with normal conductivity $ \sigma_{\perp} $ and tangential conductivity $\sigma_{\parallel}$), the only component $ \sigma_{\perp} $ is relevant for the low conducting model and the only component $\sigma_{\parallel}$ is relevant for the high conducting interface.

Of course, if both relations $g=0$ and $r^-=r^+=0$ are satisfied in the T-model, then the ideal interface is simply obtained. 
It is not difficult to prove that this model is completely equivalent to a series of three different ideal sheets (multi-layered interface) A, B and C: an external low conducting phase with Kapitza resistance $r^+=\lim_{\Delta h^{(A)} \rightarrow 0, \sigma_{\perp}^{(A)} \rightarrow 0}\frac{\Delta h^{(A)}}{\sigma_{\perp}^{(A)}}$, a halfway high conducting phase with specific conductance $g=\lim_{\Delta h^{(B)} \rightarrow 0, \sigma_{\parallel}^{(B)} \rightarrow \infty}  \sigma_{\parallel}^{(B)}\Delta h^{(B)}$ and, finally, an internal low conducting phase with Kapitza resistance $r^-=\lim_{\Delta h^{(C)} \rightarrow 0, \sigma_{\perp}^{(C)} \rightarrow 0}\frac{\Delta h^{(C)}}{\sigma_{\perp}^{(C)}}$. The three layers are characterized by thickness $\Delta h^{(A)}$, $\Delta h^{(B)}$, $\Delta h^{(C)}$ (with $\Delta h=\Delta h^{(A)}+\Delta h^{(B)}+\Delta h^{(C)}$) and conductivities $\sigma_{\perp}^{(A)}$, $\sigma_{\parallel}^{(B)}$, $\sigma_{\perp}^{(C)}$. So, we have built an example of interpretation of the model parameters with a concrete physical multilayered structure. 

A dual model can be introduced by considering the second structure depicted in Fig.\ref{figure1} ($\Pi$-model). A procedure similar to the previous one leads to the following interface equations
\begin{eqnarray}
\label{saltoV2}
\llbracket V \rrbracket &=& -r \left( \vec{J}\cdot\vec{n}\right)^+ +rg^+\frac{\partial^{2}}{\partial s^{2}}V^{+},
\\
\label{saltoJ2}
\llbracket \vec{J}\cdot\vec{n} \rrbracket &=& g^+\frac{\partial^{2}}{\partial s^{2}}V^+ +g^-\frac{\partial^{2}}{\partial s^{2}}V^{-},
\end{eqnarray}
where the parameters $ r $, $g^+$ and $g^-$ are the suitably rescaled counterparts of $R$, $G^+$ and $G^-$, appearing in Fig.\ref{figure1}, right. As before, the operator $\partial^2/\partial s^2$ must be substituted with the surface Laplacian $\nabla_{S}^2$ for the 3D case. 
We can prove that also the $\Pi$-model is exactly equivalent to a series of three different ideal sheets: an external high conducting phase with conductance $g^+=\lim_{\Delta h^{(A)} \rightarrow 0, \sigma_{\parallel}^{(A)} \rightarrow \infty}  \sigma_{\parallel}^{(A)}\Delta h^{(A)}$, a halfway low conducting phase with Kapitza resistance $r=\lim_{\Delta h^{(B)} \rightarrow 0, \sigma_{\perp}^{(B)} \rightarrow 0}\frac{\Delta h^{(B)}}{\sigma_{\perp}^{(B)}}$ and, finally, an internal high conducting phase with conductivity $g^-=\lim_{\Delta h^{(C)} \rightarrow 0, \sigma_{\parallel}^{(C)} \rightarrow \infty}  \sigma_{\parallel}^{(C)}\Delta h^{(C)}$. As before, the layers are characterized by thickness $\Delta h^{(A)}$, $\Delta h^{(B)}$, $\Delta h^{(C)}$  and conductivities $\sigma_{\parallel}^{(A)}$, $\sigma_{\perp}^{(B)}$, $\sigma_{\parallel}^{(C)}$. It is important to remark that the interpretation of the model through three adjacent layers (for both the T and $\Pi$ structures) it is not restrictive; in fact, the proposed schemes can be also used to effectively represent different imperfect interfaces with all parameters fitted in order to mimic their correct behavior. We also underline that some more complete models have been proposed in literature (see, e.g., the recent Gu and He interface,\cite{gu-he} which also degenerates to the high or low conducting models and it is able to take into account all the coupling among the electric, magnetic and elastic fields); here, we have proposed our schemes with the idea to find a compromise between the complexity and the possibility to analytically solve the problem for paradigmatic composite structures. 

The proposed models are dual from both the geometrical point of view, as shown by the T and $\Pi$ lattice structures, and the physical point of view, as discussed below through the results for the composite materials.

\section{Single particle behavior}

\begin{figure}[t!]
\includegraphics[width=8.2cm]{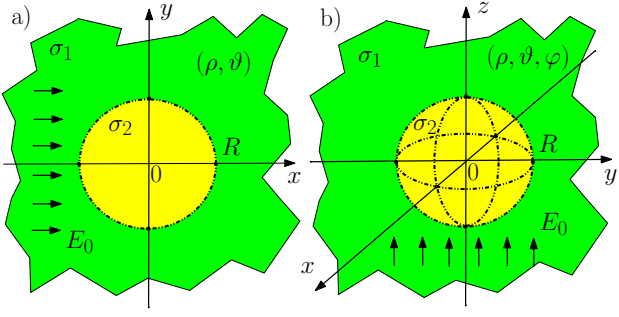}
\caption{\label{figure2} (Color online) Scheme of a single circular (a) or spherical (b) particle with conductivity $\sigma_2$ embedded into a matrix with conductivity $\sigma_1$. The interface between the two phases is described by either Eqs.(\ref{saltoV1}) and (\ref{saltoJ1}) or Eqs.(\ref{saltoV2}) and (\ref{saltoJ2}).}
\end{figure}

We consider now a single circular (in 2D) or spherical (in 3D) particle with conductivity $\sigma_2$ embedded into a matrix with conductivity $\sigma_1$ (see Fig.\ref{figure2}): we suppose that the interface between the constituents is described by Eqs.(\ref{saltoV1}) and (\ref{saltoJ1}) (T-model) and we determine the effect of an arbitrary externally applied field. 
Since we are dealing with two isotropic phases, in order to solve the problem we can directly apply the original idea of Maxwell,\cite{maxwell} which is based on the following steps. 
Firstly, we observe that the electrical potential must be an harmonic function both inside and outside the particle: therefore, it can be straightforwardly expanded in trigonometric series (in 2D) and in series of spherical harmonics (in 3D).\cite{prospe}
Secondly, we can substitute such expansions in the interface conditions, by obtaining a set of equations for the unknown coefficients, completely describing the potential both inside and outside the inhomogeneity.
To accomplish this last step we must determine the surface Laplacian of the series expansions: to do this we remember that the trigonometric functions and the spherical harmonics are eigenfunctions of the $\nabla_{S}^{2}$ operator with certain eigenvalues described in Appendix A. 
The complete procedure, which is valid for any externally applied field, is described in Appendix B for the 2D case and in Appendix C for the 3D case. Here, we are interested in the particular case with an uniform applied electric field $E_0$, corresponding to a potential $V_0=-\rho \cos \vartheta E_0$ (see Fig.\ref{figure2}). 
The perturbation induced by the inhomogeneity with imperfect contact has been eventually found as
\begin{eqnarray}
\label{dentro}
&&\mbox{for  } \rho<R \Rightarrow V=-\rho \cos \vartheta E_{0}\left( \frac{d\sigma_1 }{\mathcal{C}} \right),  \\
\label{fuori}
&&\mbox{for  } \rho>R \Rightarrow V=-\rho \cos \vartheta E_{0}\left(  1+\frac{R^{d}}{\rho^d}\frac{\mathcal{B}}{\mathcal{C}}\right) , 
\end{eqnarray}
where $d=2$ for the circle, $d=3$ for the sphere and the parameters $\mathcal{B}$ and $\mathcal{C}$ are defined as follows
\begin{eqnarray}
\nonumber
\mathcal{B}&=&\sigma_{1}-\sigma_{2}+\frac{r^+ +r^-}{R}\sigma_{1}\sigma_{2}\\
\label{bbb}
&&-(d-1)\frac{g}{R}\left[1-r^+\frac{\sigma_{1}}{R}\right] \left[ 1+ r^-\frac{\sigma_{2}}{R}\right], \\
\nonumber
\mathcal{C}&=&(d-1)\sigma_{1}+\sigma_{2}+(d-1)\frac{r^+ +r^-}{R}\sigma_{1}\sigma_{2}\\
\label{ccc}
&&+(d-1)\frac{g}{R}\left[1+ (d-1)r^+\frac{\sigma_{1}}{R}\right] \left[ 1+ r^-\frac{\sigma_{2}}{R}\right].
\end{eqnarray}
We can observe that the electric quantities both inside and outside the particle, in contrast to the case with perfect interfaces, depend on $R$. So, the previous result can be used for analysing the scale effects induced by the imperfect contact. 
From Eq.(\ref{dentro}) it is easy to identify the induced internal field as $E_{int}/E_0={d\sigma_1 }/{\mathcal{C}}$. A first scaling law for $R\rightarrow \infty$ can be obtained by introducing the classical Lorentz field for a particle with a perfect interface $E_{lor}=E_{int}\mid_{r^+=r^-=0,g=0}$; we can easily prove that
\begin{eqnarray}
\nonumber
\frac{E_{int}}{E_{lor}}-1=-\frac{(d-1)\sigma_{1}}{(d-1)\sigma_{1}+\sigma_{2}}\frac{\ell^- +\ell^+ +\mathcal{L}}{R}+O\left(\frac{1}{R^{2}} \right), \\
\label{scalefield}
\end{eqnarray}
where we have introduced the following intrinsic length scales
\begin{eqnarray}
\ell^-=\sigma_{2}r^-,\,\,\,\ell^+=\sigma_{2}r^+,\,\,\,\mathcal{L}=\frac{g}{\sigma_1},
\label{lengths}
\end{eqnarray}
 which automatically emerge from the analysis and completely control all the scaling laws.
Eq.(\ref{scalefield}) means that the internal field approaches the Lorentz field for large radius of the particle ($R\gg\ell^- +\ell^+ +\mathcal{L}$), i.e. the effects of the contact imperfection are vanishingly small for $R\rightarrow \infty$.

\begin{figure}[t]
\includegraphics[width=8.5cm]{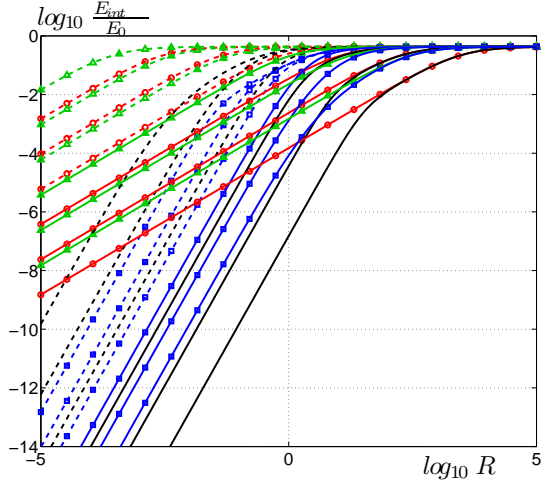}
\caption{\label{figure3}(Color online) Plot of $log_{10}(E_{int}/E_0)$ versus $log_{10}R$ for the T-model. Green curves with triangles: high conductivity model with a varying $g$ in $\Omega=\left\lbrace 0.001, 0.016, 0.25, 4, 64, 1000 \right\rbrace $. Red curves with circles: low conductivity model with a varying $r^+=r^-$ in $\Omega$. Blue curves with squares: general model with $r^+=r^-=1$ and $g$ varying in $\Omega$. Black curves without symbols: general model with $g=1$ and $r^+=r^-$ varying in $\Omega$. Everywhere, the dashed lines correspond to the values $<1$ of the varying quantity. Parameters adopted in a.u.: $\sigma_1=1$, $\sigma_2=5$, $d=3$, $c=0.3$.}
\end{figure}

We discuss now the scaling laws obtained for $R\rightarrow 0$. A long but straightforward analysis leads to
\begin{eqnarray}
\label{fieldint}
\frac{E_{int}}{E_0}&=&\frac{d\sigma_2}{(d-1)^{2}\sigma_1}\frac{R^{3}}{\ell^- \ell^+ \mathcal{L}}+O\left({R^{4}} \right), \\
\label{fieldlow}
\left.\frac{E_{int}}{E_0}\right\vert_{g=0}&=&\frac{d}{(d-1)}\frac{R}{\ell^- +\ell^+}+O\left({R^{2}} \right), \\
\label{fieldhigh}
\left.\frac{E_{int}}{E_0}\right\vert_{r^+=r^-=0}&=&\frac{d}{(d-1)}\frac{R}{\mathcal{L}}+O\left({R^{2}} \right).
\end{eqnarray}
In any case the internal field converges to zero for very small particles.
It is interesting to observe that the internal field for the T-model follows a scaling law with a power of three, while the low and high conductivity models follow a law with a scaling exponent equal to one. 
It can be seen in Fig.\ref{figure3} where $log_{10}(E_{int}/E_0)$ is represented versus $log_{10}R$. 
The blue curves (with squares) and the black ones (without symbols) describe the generic interface ($g\neq 0$, $r^+\neq 0$, $r^-\neq 0$) and show a slope +3 for small $R$, which is in agreement with Eq.(\ref{fieldint}). 
On the other hand, green curves (with triangles) and red ones (with circles) correspond to the high and the low conductivity interface, respectively: they all exhibit a slope +1 for small $R$ as predicted by Eqs.(\ref{fieldlow}) and (\ref{fieldhigh}). 
We also note that all curves in Fig.\ref{figure3} converge to the Lorentz field for $R\rightarrow \infty$, as described by Eq.(\ref{scalefield}).

Now, we take into consideration the $\Pi$-model described by Eqs.(\ref{saltoV2}) and (\ref{saltoJ2}). The perturbation to the electric potential generated by the inhomogeneity is described again by Eqs.(\ref{dentro}) and (\ref{fuori}) but with new coefficients $\mathcal{B}$ and $\mathcal{C}$ given below 
\begin{eqnarray}
\nonumber
\mathcal{B}&=&\sigma_{1}-\sigma_{2}+\frac{r}{R}\sigma_{1}\sigma_{2}-(d-1)^2\frac{g^- g^+ r}{R^3}\\
\label{bbbbis}
&&-\frac{(d-1)}{R}\left\lbrace g^-\left[1-r\frac{\sigma_{1}}{R}\right]+g^+ \left[ 1+ r\frac{\sigma_{2}}{R}\right]\right\rbrace , \\
\nonumber
\mathcal{C}&=&(d-1)\sigma_{1}+\sigma_{2}+(d-1)\frac{r}{R}\sigma_{1}\sigma_{2}+(d-1)^2\frac{g^- g^+ r}{R^3}\\
\nonumber
&&+\frac{(d-1)}{R}\left\lbrace g^-\left[1+ (d-1)r\frac{\sigma_{1}}{R}\right]+g^+ \left[ 1+ r\frac{\sigma_{2}}{R}\right]\right\rbrace . \\
\label{cccbis}
\end{eqnarray}
With regards to the scaling law for $R\rightarrow\infty$, it is possible to prove that Eq.(\ref{scalefield}) must be substituted with
\begin{eqnarray}
\nonumber
\frac{E_{int}}{E_{lor}}-1=-\frac{(d-1)\sigma_{1}}{(d-1)\sigma_{1}+\sigma_{2}}\frac{\ell +\mathcal{L}^+ +\mathcal{L}^-}{R}+O\left(\frac{1}{R^{2}} \right), \\
\label{scalefield2}
\end{eqnarray}
where we have introduced the dual intrinsic length scales
\begin{eqnarray}
\ell=\sigma_{2}r,\,\,\,\mathcal{L}^+=\frac{g^+}{\sigma_1},\,\,\,\mathcal{L}^-=\frac{g^-}{\sigma_1}.
\label{lengthsdual}
\end{eqnarray}
As expected, also in this case the internal field approaches the Lorentz field for large radius of the particle ($R\gg\ell +\mathcal{L}^+ +\mathcal{L}^-$).
On the other hand, for $R\rightarrow 0$, Eqs.(\ref{fieldint})-(\ref{fieldhigh}) become as follows
\begin{eqnarray}
\label{fieldint1}
\frac{E_{int}}{E_0}&=&\frac{d\sigma_2}{(d-1)^{2}\sigma_1}\frac{R^{3}}{\ell \mathcal{L}^+\mathcal{L}^-}+O\left({R^{4}} \right), \\
\label{fieldlow1}
\left.\frac{E_{int}}{E_0}\right\vert_{g^+=g^-=0}&=&\frac{d}{(d-1)}\frac{R}{\ell}+O\left({R^{2}} \right), \\
\label{fieldhigh1}
\left.\frac{E_{int}}{E_0}\right\vert_{r=0}&=&\frac{d}{(d-1)}\frac{R}{\mathcal{L}^++\mathcal{L}^-}+O\left({R^{2}} \right).
\end{eqnarray}
As before, the internal field converges to zero for very small particles (with different scaling exponents, as above described).

\section{Effective conductivity of dispersions}

To analyse the effects of imperfect interfaces on a composite material we consider a dispersion of cylindrical or spherical particles of conductivity $\sigma_2$ in a matrix with conductivity $\sigma_1$. 
When the interfaces are described by Eqs.(\ref{saltoV1}) and (\ref{saltoJ1}) (T-model) we can generalize the Maxwell approach\cite{maxwell} or, equivalently, the Mori-Tanaka scheme\cite{mori} by obtaining the following effective conductivity for a composite with a volume fraction $c$ of the dispersed particles 
\begin{eqnarray}
\label{maxgen}
\dfrac{\sigma_{eff}}{\sigma_{1}}=\frac{1}{ 1+\dfrac{cd\mathcal {B} }{(1-c)\mathcal{C} +c \left[ \mathcal{C}-(d-1)\mathcal {B} \right]}},
\end{eqnarray}
where $\mathcal {B} $ and $\mathcal{C} $ are given in Eqs.(\ref{bbb}) and (\ref{ccc}). 
Detailed descriptions of the homogenization procedures can be found elsewhere.\cite{bonnet,bonnet-he,giorda, giorda1}
For interfaces described by the low conductivity model we obtain $\sigma_{low}=\sigma_{eff}\mid_{g=0}$, which is in perfect agreement with recent investigations;\cite{bonnet,duan-thermal} on the other hand, when the high conductivity model is accounted for we have  $\sigma_{high}=\sigma_{eff}\mid_{r^+=r^-=0}$, which corresponds to some known results.\cite{bonnet-he,duan-thermal}
Moreover, when we consider a perfect contact between the constituents we obtain the celebrated Maxwell formula\cite{maxwell}
\begin{eqnarray}
\dfrac{\sigma_{max}}{\sigma_{1}}=\frac{1}{1+\dfrac{dc(\sigma_{1}-\sigma_{2})}{(1-c)\left[(d-1)\sigma_{1}+\sigma_{2} \right] +cd\sigma_{2}} }.
\end{eqnarray}
The first important scaling law concerns the situation with a large radius of the particles: in this case, as above said, the size effects disappear and the effective conductivity converges to the Maxwell one as follows
\begin{eqnarray}
\label{largeR}
\dfrac{\sigma_{eff}}{\sigma_{max}}-1=\frac{cd^2 \sigma_1^{2}  }{\mathcal{G} }\frac{\mathcal{H}}{R}+O\left(\frac{1}{R^{2}} \right),
\end{eqnarray}
where we have defined
\begin{eqnarray}
\mathcal{H}&=&(d-1)\mathcal{L}-\frac{\sigma_{2}}{\sigma_{1}} (\ell^+ +\ell^-),\\
\nonumber
\mathcal{G}&=&\left[(d+c-1)\sigma_{1}+(1-c)\sigma_{2} \right]\\
&&\times \left[(d-1)(1-c)\sigma_{1}+(cd-c+1)\sigma_{2} \right].
\label{GGG}
\end{eqnarray}
The parameter $\mathcal{H}$ represents the overall length scale of this process and it is a linear combination of the terms defined in Eq.(\ref{lengths}). This result can be simply compared with recent achievements\cite{duan-thermal} concerning the  cases with low and high conductivity interfaces. Indeed, we find a perfect agreement if $\ell^+ =\ell^-=0$ or $\mathcal{L}=0$.

Other interesting scaling laws can be found for $R\rightarrow 0$. To analyse this case we define the conductivity $\sigma_{0}$, which represents a Maxwell dispersion with $\sigma_2 \rightarrow 0$ (dispersion of voids), and the conductivity $\sigma_{\infty}$, which characterizes a Maxwell dispersion with $\sigma_2 \rightarrow \infty$ (dispersion of superconducting particles):
\begin{eqnarray}
\dfrac{\sigma_{0}}{\sigma_{1}}=\frac{(1-c)(d-1)}{d+c- 1 };\,\,  \dfrac{\sigma_{\infty}}{\sigma_{1}}=\frac{1-c+cd}{1-c }.
\end{eqnarray}
First of all we observe that if $r^+ \neq 0$ we have ${\sigma_{eff}\rightarrow}{\sigma_{0}}$ with  the scaling law
\begin{eqnarray}
\nonumber
\dfrac{\sigma_{eff}}{\sigma_{0}}-1=\frac{cd^2 \sigma_2}{(d-1)(1-c)(d-1+c)\sigma_1}\frac{R}{\ell^+}+O\left({R}^2 \right).\\
\end{eqnarray}
Similarly, if $r^- \neq 0$ with $r^+ = 0$ and $g=0$ we obtain ${\sigma_{eff}\rightarrow}{\sigma_{0}}$ with the scaling law
\begin{eqnarray}
\nonumber
\dfrac{\sigma_{eff}}{\sigma_{0}}-1=\frac{cd^2 \sigma_2}{(d-1)(1-c)(d-1+c)\sigma_1}\frac{R}{\ell^-}+O\left({R}^2 \right).\\
\end{eqnarray}
So, for the general T-model and for the low conducting interface we have ${\sigma_{eff}\rightarrow}{\sigma_{0}}$ (when $R\rightarrow 0$) with a scaling exponent equals to one.
On the other hand, for $g\neq 0$ and $r^+=0$ we prove the convergence ${\sigma_{eff}\rightarrow}{\sigma_{\infty}}$ with a scaling law
\begin{eqnarray}
\nonumber
\dfrac{\sigma_{eff}}{\sigma_{\infty}}-1=-\frac{cd^2 }{(d-1)(1-c)(cd-c+1)}\frac{R}{\mathcal{L}}+O\left({R}^2 \right).\\
\end{eqnarray}
It means that the high conductivity model leads to ${\sigma_{eff}\rightarrow}{\sigma_{\infty}}$ for $R\rightarrow 0$.

\begin{figure}[t]
\includegraphics[width=8.5cm]{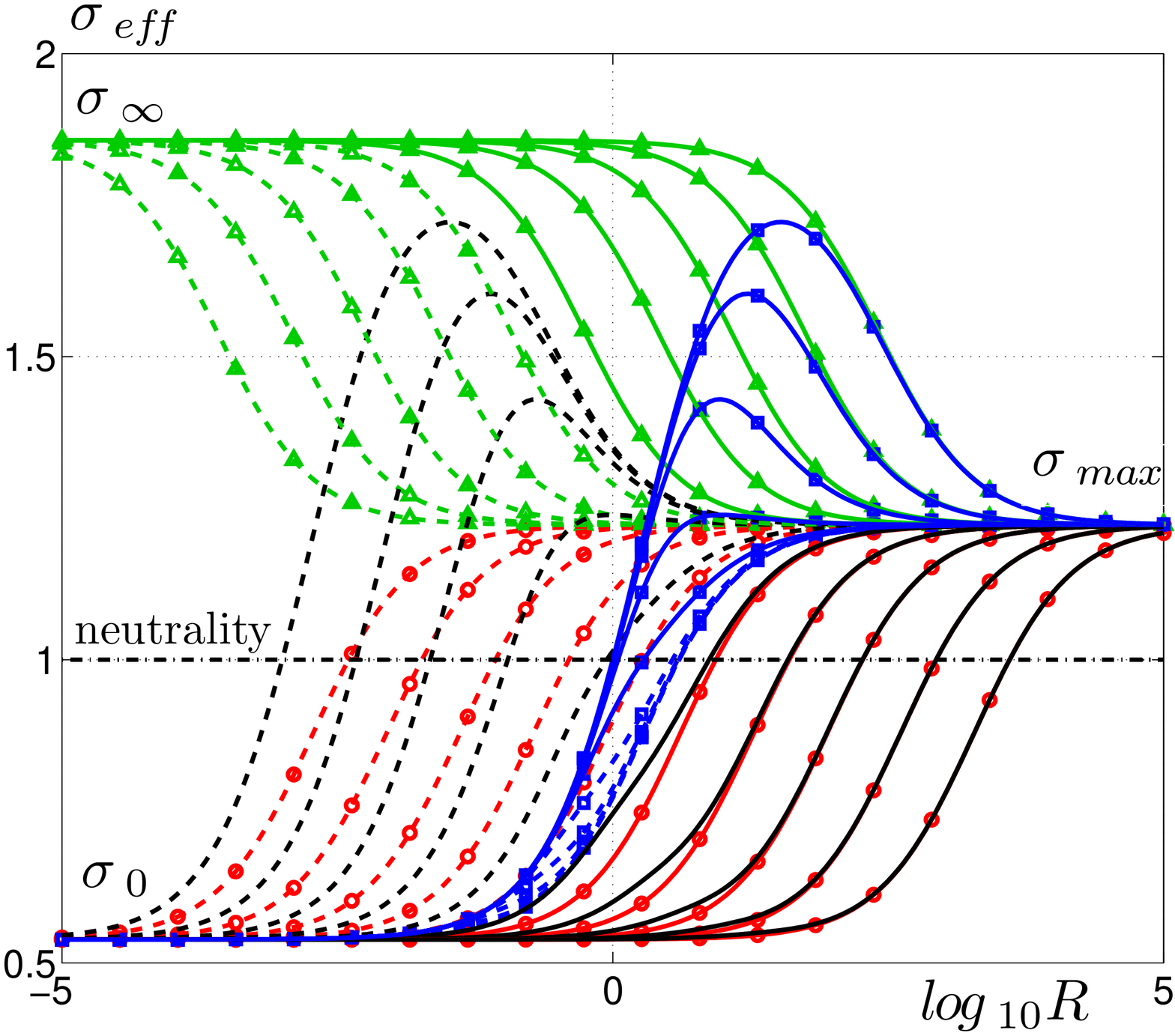}
\includegraphics[width=8.5cm]{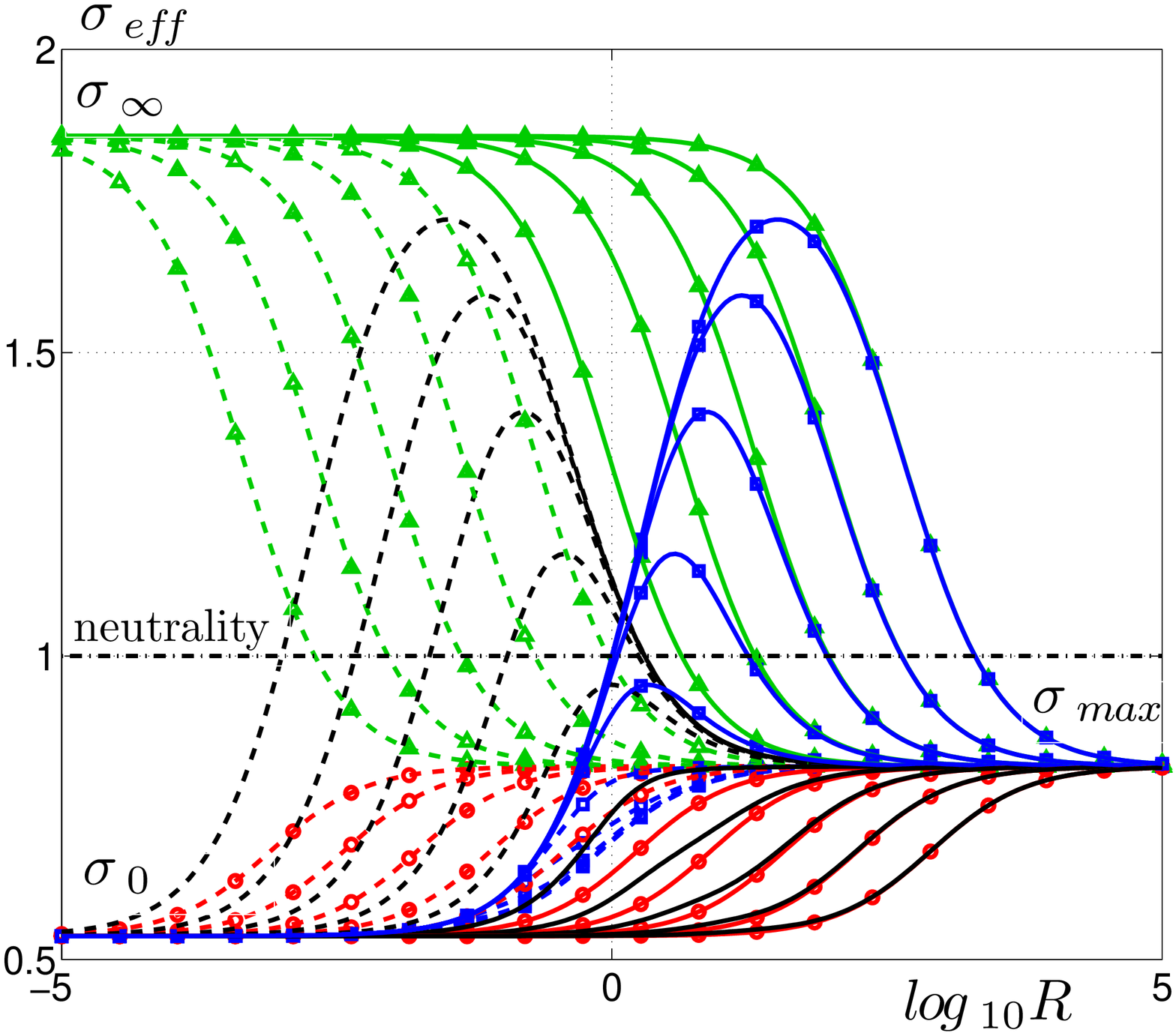}
\caption{\label{figure4}(Color online) Plot of $\sigma_{eff}$ versus $log_{10}R$ for the T-model. We adopted the following parameters (in a.u.): $\sigma_1=1$, $\sigma_2=2$ (top) and $\sigma_1=1$, $\sigma_2=1/2$ (bottom). Everywhere we used $d=2$, $c=0.3$. Green curves with triangles: high conductivity model ($ r^+=r^-=0$) with a varying $g$ in $\Omega=\left\lbrace 10^{-3+2(j-1)/3}, j=1...10 \right\rbrace $. Red curves with circles: low conductivity model ($g=0$) with a varying $r^+=r^-$ in $\Omega$. Blue curves with squares: T-model with $r^+=r^-=1$ and $g$ varying in $\Omega$. Black curves without symbols: T-model with $g=1$ and $r^+=r^-$ varying in $\Omega$. Everywhere, the dashed lines correspond to values $<1$ of the varying quantity.
}
\end{figure}

\begin{figure}[t]
\includegraphics[width=8.5cm]{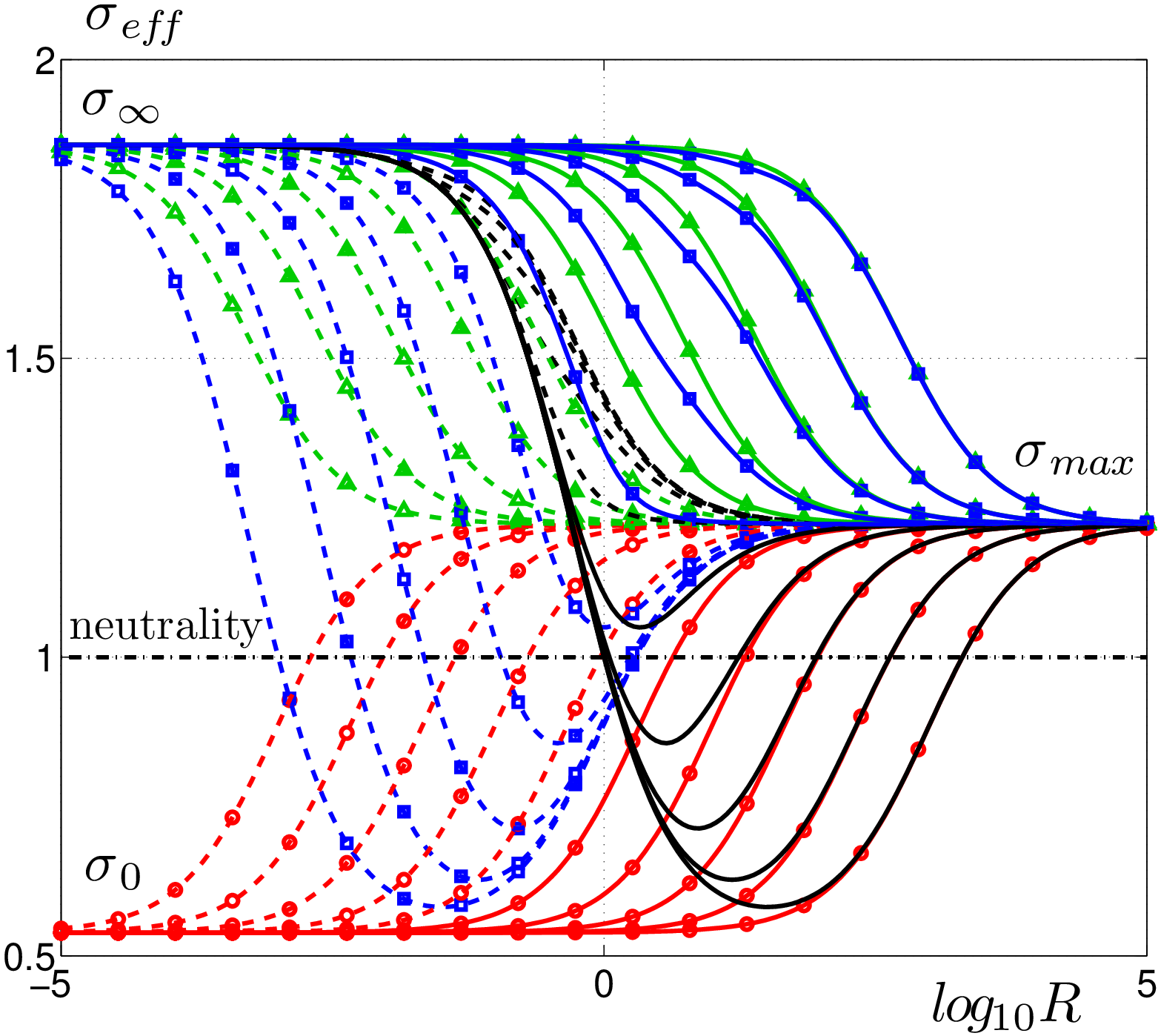}
\includegraphics[width=8.5cm]{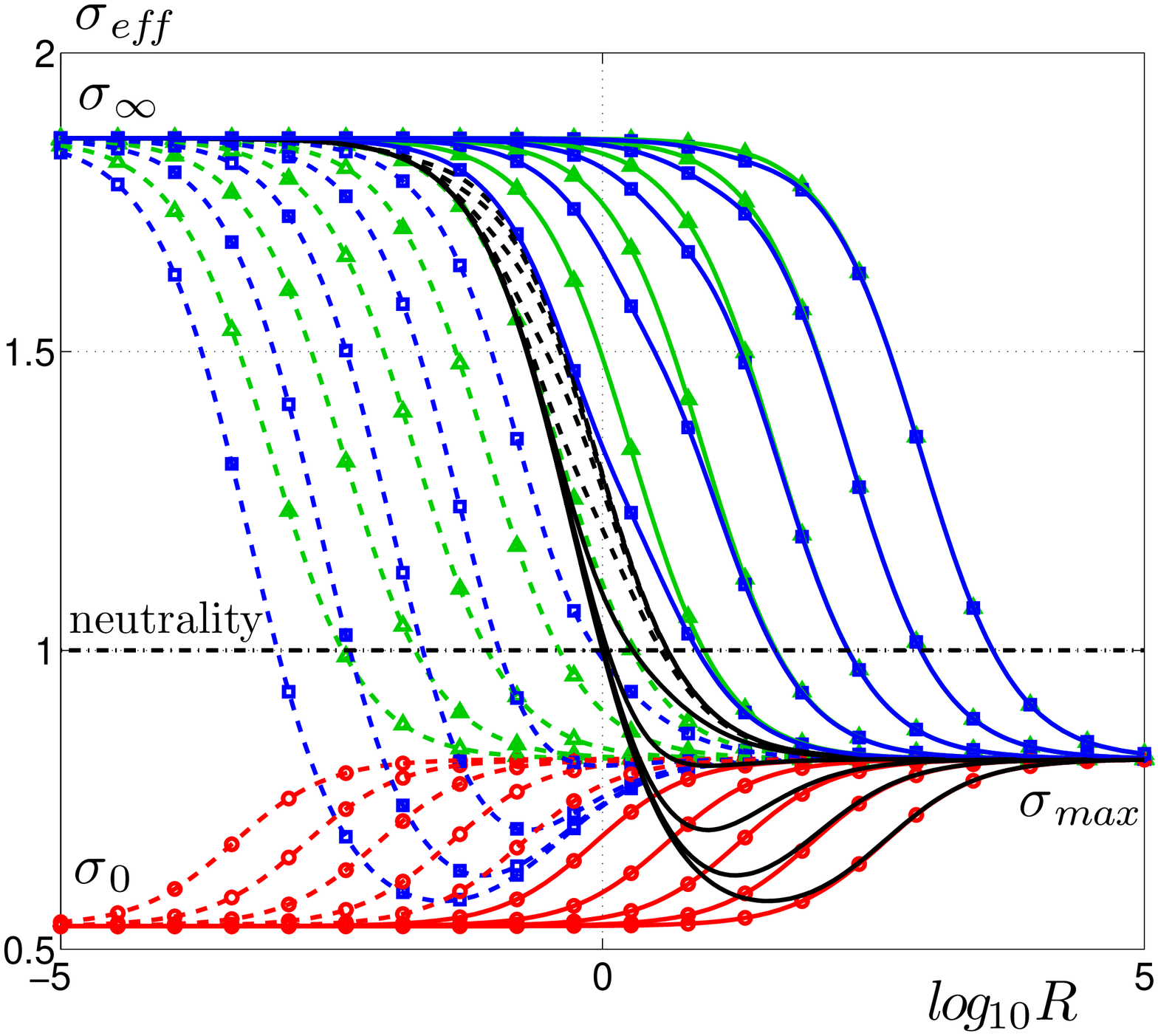}
\caption{\label{figure5}(Color online) Plot of $\sigma_{eff}$ versus $log_{10}R$ for the $\Pi$-model. We adopted the following parameters (in a.u.): $\sigma_1=1$, $\sigma_2=2$ (top) and $\sigma_1=1$, $\sigma_2=1/2$ (bottom). Everywhere we used $d=2$, $c=0.3$. Green curves with triangles: high conductivity model ($r=0$) with a varying $g^-=g^+$ in $\Omega=\left\lbrace 10^{-3+2(j-1)/3}, j=1...10 \right\rbrace $. Red curves with circles: low conductivity model ($g^+=g^-=0$) with a varying $r$ in $\Omega$. Blue curves with squares: $\Pi$-model with $r=1$ and $g^+=g^-$ varying in $\Omega$. Black curves without symbols: $\Pi$-model with $g^+=g^-=1$ and $r$ varying in $\Omega$. Everywhere, the dashed lines correspond to values $<1$ of the varying quantity. 
}
\end{figure}

This complex scenario is summarized in Fig.\ref{figure4} where $\sigma_{eff}$ is shown versus $log_{10}R$. 
Even at constant volume fraction $c$, significant size effects on the effective conductivity are evident for a variable radius $R$.
In Fig.\ref{figure4} (top) we have reported the results for $\sigma_2/\sigma_1=2$ and in Fig.\ref{figure4} (bottom) for $\sigma_2/\sigma_1=0.5$. 
In both cases we have shown the neutrality axis at which the effective conductivity $\sigma_{eff}$ equals the matrix conductivity $\sigma_1$, making the inclusions effectively hidden.\cite{tor-rin, benmiloneutrality} 
We can observe that $\sigma_{eff}$ is a monotonically decreasing function of $R$ (from $\sigma_{\infty}$ to $\sigma_{max}$) for the high conductivity model (green curves with triangles), while it is a monotonically increasing function of $R$ (from $\sigma_{0}$ to $\sigma_{max}$) for the low conductivity model (red lines with circles). 
So, the neutrality condition ($\mathcal{B}=0$) can be satisfied by the low conductivity model for $\sigma_2>\sigma_1$ ($\sigma_2-\sigma_1=r\sigma_1\sigma_2/R$, see Fig.\ref{figure4}, top) and by the high conductivity model for $\sigma_2<\sigma_1$ ($\sigma_1-\sigma_2=g(d-1)/R$, see Fig.\ref{figure4}, bottom). 
On the other hand, the blue and black lines concern the case of the general T-model and they exhibit a non monotone behavior starting from $\sigma_{0}$ and arriving at $\sigma_{max}$. 
It is interesting to note that, with the general T-model, it is possible to satisfy the neutrality condition for both the cases $\sigma_2>\sigma_1$ and $\sigma_2<\sigma_1$. The condition leading to neutrality in this case (T-model) is
\begin{eqnarray}
\label{neutt}
g&=&\frac{\sigma_{1}-\sigma_{2}+\frac{r^+ +r^-}{R}\sigma_{1}\sigma_{2}}{(d-1)\frac{1}{R}\left[1-r^+\frac{\sigma_{1}}{R}\right] \left[ 1+ r^-\frac{\sigma_{2}}{R}\right] },
\end{eqnarray}
which is represented in Fig.\ref{figure4} by the intersections of blue and black curves with the neutrality axis.

As for the dual $\Pi$-model, we can affirm that the generalized Maxwell theory given in Eq.(\ref{maxgen}) is still valid but the coefficients  $\mathcal{B}$ and $\mathcal{C}$ must be taken from Eqs.(\ref{bbbbis}) and (\ref{cccbis}), respectively. For a large radius of the particle we have the scaling law identical to Eq.(\ref{largeR}) where $\mathcal{G}$ is given by Eq.(\ref{GGG}) and $\mathcal{H}$ by the following expression
\begin{eqnarray}
\mathcal{H}&=&(d-1)(\mathcal{L}^+ +\mathcal{L}^-)-\frac{\sigma_{2}}{\sigma_{1}}\ell .
\end{eqnarray}
It represents the overall length scale of the $\Pi$-model and it is indeed a linear combination of the terms defined in Eq.(\ref{lengthsdual}). 
We also report the scaling laws for $R\rightarrow 0$. 
If $g^+ \neq 0$ we have that ${\sigma_{eff}\rightarrow}{\sigma_{\infty}}$ with  the scaling law
\begin{eqnarray}
\nonumber
\dfrac{\sigma_{eff}}{\sigma_{\infty}}-1=-\frac{cd^2 }{(d-1)(1-c)(cd-c+1)}\frac{R}{\mathcal{L}^+}+O\left({R}^2 \right).\\
\end{eqnarray}
Similarly, if $g^- \neq 0$ with $r = 0$ and $g^+=0$ we obtain ${\sigma_{eff}\rightarrow}{\sigma_{\infty}}$ with the scaling law
\begin{eqnarray}
\nonumber
\dfrac{\sigma_{eff}}{\sigma_{\infty}}-1=-\frac{cd^2 }{(d-1)(1-c)(cd-c+1)}\frac{R}{\mathcal{L}^-}+O\left({R}^2 \right).\\
\end{eqnarray}
So, for the general $\Pi$-model and for the high conducting interface we have ${\sigma_{eff}\rightarrow}{\sigma_{\infty}}$ (when $R\rightarrow 0$) with a scaling exponent equals to one.
On the other hand, for $r\neq 0$ and $g^+=0$ we prove the convergence ${\sigma_{eff}\rightarrow}{\sigma_{0}}$ with a scaling law
\begin{eqnarray}
\nonumber
\dfrac{\sigma_{eff}}{\sigma_{0}}-1=\frac{cd^2 \sigma_2}{(d-1)(1-c)(d-1+c)\sigma_1}\frac{R}{\ell}+O\left({R}^2 \right).\\
\end{eqnarray}
It means that, as expected, the low conductivity model leads to ${\sigma_{eff}\rightarrow}{\sigma_{0}}$ for $R\rightarrow 0$.

In Fig.\ref{figure5} the results for the $\Pi$-model are shown: the effective conductivity is represented versus the radius $R$ of the particles. In Fig.\ref{figure5} (top) we have the case with $\sigma_2/\sigma_1=2$ and in Fig.\ref{figure5} (bottom) we show the results for $\sigma_2/\sigma_1=0.5$. All previous scaling laws are confirmed and clearly indicated. By drawing a comparison between  Fig.\ref{figure4} and Fig.\ref{figure5} we can point out the dual character of the proposed models: the T-model behaves similarly to the low conducting interface with regards to the limiting cases $R\rightarrow 0$ and $R \rightarrow \infty$ but it shows a specific additional upwards peak describing the competition of the scale effects with the presence of the tangential conductances $g$.
Conversely, the $\Pi$-model behaves similarly to the high conducting interface with regards to the limiting cases $R\rightarrow 0$ and $R \rightarrow \infty$ but it shows a specific additional downwards peak describing the competition of the scale effects with the presence of the normal resistances $r$. As before, also in the case of the $\Pi$-model, we can satisfy the neutrality condition for both the contrast situations $\sigma_2/\sigma_1>1$ and $\sigma_2/\sigma_1<1$. 
The condition leading to neutrality in this case ($\Pi$-model) is
\begin{eqnarray}
\label{neupi}
r&=&\frac{\sigma_{1}-\sigma_{2}-\frac{g^+ +g^-}{R}(d-1)}{(d-1)^{2}\frac{g^+ +g^-}{R^3}-\frac{d-1}{R^2}(g^-\sigma_{1}-g^+\sigma_{2})-\frac{\sigma_{1}\sigma_{2}}{R} },\,\,\,\,
\end{eqnarray}
and it is satisfied in Fig.\ref{figure5} at the intersection points between the blue or blacks curves and the neutrality axis.

By means of this analysis we can assert that the T and $\Pi$ models exhibit an interesting complex behavior which is able to reproduce many properties of real interfaces appearing in different nano-systems. As an example we can compare our results with those recently obtained for a dispersion of SiC particles (with radius between 5 and 15\AA) in a polymeric (epoxy) matrix.\cite{appl7} By means of a multiscale combination of the non-equilibrium molecular dynamics and a micromechanics bridging model, the thermal conductivity has been studied in terms of the particles radius. The  result is in perfect qualitative agreement with our T-model and a maximum value of the conductivity was obtained for a given radius. To obtain such a result the Kapitza resistance and a specific interphase describing the bonding of the polymers to the monocrystalline SiC particles have been considered.\cite{appl7} Our T-model is able to describe the overall response of the structured/multilayered interface through the simple conditions given in Eqs.(\ref{saltoV1}) and (\ref{saltoJ1}) imposing the jumps of the physical fields over the zero-thickness interface. Therefore, the proposed models perfectly implement the multiscale paradigm by introducing the effective properties of a given interface behavior.      

We remark that in this Section we have used the generalization of the Maxwell  approach\cite{maxwell} or the Mori-Tanaka scheme\cite{mori} in order to obtain simple results and to directly analyse the scale effects induced by the imperfect interfaces. Nevertheless, the closed form results discussed in Section III for the single particle response can be easily exploited to implement other homogenization techniques such as the differential method\cite{michel,giorda,epjb}, the self consistent scheme\cite{lak1,gi1,gi2}, the generalized-self-consistent model\cite{lo} and the strong-property-fluctuation theory.\cite{mackay} We also remark that the analysis of the imperfect interfaces is an important topic also in the field of micromechanics (elasticity of composites) where several theoretical models have been proposed\cite{jia1,jia2,kar} and intriguing scale effects have been observed.\cite{kar,palla,ropp}

\section{Summary and conclusions}
In this paper we have taken into consideration the possible scale effects induced by imperfect interfaces between the constituents of an heterogeneous system. To this aim we introduced two generalised schemes, namely the T and $\Pi$ structures, which can be seen as natural combinations of the so-called low and high conducting interface models. 
One important property discussed concerns the uniformity of the physical fields in circular or spherical particles with T or $\Pi$ imperfect interfaces. 
This point extends well known theorems proving the uniformity in different conditions and opens the possibility to study the behavior of new interfaces in anisotropic, elliptic and ellipsoidal particles, which are standard problems in the theory of inhomogeneities. 
The results for a single inclusion were applied to the analysis of the effective properties of dispersions. In particular we studied the scale effects and we found  interesting behaviors, which generalize those observed with low and high conductivity interfaces. 
We indeed observed a specific peak of the effective conductivity in correspondence to a critical radius of the dispersed particles: it corresponds to the competition between the tendency to attain the Maxwell conductivity limit for a large radius and the conduction properties of the interface, which tend to increase or decrease the overall conductivity, depending on the specific parameters. 
This is exactly the trend observed in recent analysis of imperfect interfaces in nanocomposites (hard particles in polymeric matrix or similar mixtures). 
To conclude, we have analysed the neutrality properties of the T and $\Pi$ models: contrarily to the low and high conducting interface, we have proved that it is possible to satisfy the neutrality condition for any contrast $\sigma_2/\sigma_1$ between the conductivities of the involved phases. 
So, Eqs.(\ref{neutt}) and (\ref{neupi}) are the updated versions of the  neutrality criteria, representing the generalizations of some findings, published in recent literature.
We remark that all the achievements of the present paper can be also used in dynamic regime if we consider a wavelength $\lambda$ of the propagating waves that is much larger than the radius $R$ of the particles. In this case we are working in the so-called quasi-static regime and any inhomogeneity feels a nearly static applied field.

\appendix

\section{The surface Laplacian}
The surface Laplacian operator is defined as
\begin{eqnarray}
\label{nablas}
\nabla_{S}^{2}f=\frac{1}{\sqrt{g}}\frac{\partial}{\partial \alpha_{i}}\left\lbrace \sqrt{g} g^{ij}\frac{\partial f}{\partial \alpha_{j}}\right\rbrace ,
\end{eqnarray}
where $g_{ij}$ are the components of the metric tensor (the first fundamental form) of the Riemannian manifold (the surface) $\vec{r}=\vec{r}(\alpha_{1},\alpha_{2})$.\cite{carmo} It means that $g_{ij}=\frac{\partial \vec{r}}{\partial \alpha_{i}}\cdot\frac{\partial \vec{r}}{\partial \alpha_{j}}$ and the dual components $g^{ij}$ are obtained by inverting the matrix $g_{ij}$. The quantity $g$ is the determinant of $g_{ij}$. Typically, in differential geometry of two-dimensional surfaces we adopt the symbols $g_{11}=E$, $g_{12}=g_{21}=F$ and $g_{22}=G$; so, for an orthogonal system of coordinate  lines $F=0$ and Eq.(\ref{nablas}) reduces to
\begin{eqnarray}
\nonumber
\nabla_{S}^{2}f=\frac{1}{\sqrt{EG}}\left\lbrace\frac{\partial}{\partial \alpha_{1}} \left[ \sqrt{\frac{G}{E}}\frac{\partial f}{\partial \alpha_{1}}\right] +\frac{\partial}{\partial \alpha_{2}} \left[ \sqrt{\frac{E}{G}}\frac{\partial f}{\partial \alpha_{2}}\right]\right\rbrace .\\
\label{nablasEG}
\end{eqnarray}

For a planar circle $\vec{r}=(R\cos\vartheta$, $R\sin\vartheta)$ we simply have 
\begin{eqnarray}
\label{nablascyl}
\nabla_{S}^{2}f=\frac{\partial^2 f}{\partial s^2} =\frac{1}{R^2}\frac{\partial^2 f}{\partial \vartheta^2},
\end{eqnarray}
and the following property is evident
\begin{eqnarray}
\label{autocyl}
\nabla_{S}^{2}{e}^{in\vartheta}=-\frac{1}{R^2}n^2{e}^{in\vartheta}.
\end{eqnarray}
It means that the trigonometric functions $\cos n\vartheta$ and $\sin n\vartheta$ are eigenfunctions of the Laplacian operator with eigenvalues $-\frac{1}{R^2}n^2$.

For a spherical surface $\vec{r}=(R\cos\varphi\sin\vartheta$, $R\sin\varphi\sin\vartheta$, $R\cos\vartheta)$
it is possible to obtain 
\begin{eqnarray}
\nabla_{S}^{2}f=\frac{1}{R^2}\left\lbrace \frac{1}{\sin\vartheta}\frac{\partial}{\partial \vartheta}\left[\sin\vartheta\frac{\partial f}{\partial \vartheta} \right] +\frac{1}{\sin^{2}\vartheta}\frac{\partial^2 f}{\partial \varphi^2} \right\rbrace ,
\,\,\,\,\label{nablassph}
\end{eqnarray}
and we can prove that
 \begin{eqnarray}
\label{autosph}
\nabla_{S}^{2}Y_{nm}(\vartheta,\varphi)=-\frac{1}{R^2}n(n+1)Y_{nm}(\vartheta,\varphi).
\end{eqnarray}
It means that the spherical harmonics $Y_{nm}(\vartheta,\varphi)$ are eigenfunctions of the surface Laplacian operator with eigenvalues $ -\frac{1}{R^2}n(n+1) $.\cite{prospe} They are defined (for $n\geq0$, $-n\leq m \leq n$) as\cite{grad,abra}
 \begin{eqnarray}
\label{harms}
Y_{nm}(\vartheta,\varphi)=\sqrt{\frac{2n+1}{4\pi}\frac{(n-m)!}{(n+m)!}}P_{n}^{m}(\cos \vartheta){e}^{im\varphi},
\end{eqnarray}
where $P_{n}^{m}(\xi)$ are the associated Legendre polynomials\cite{grad,abra}
 \begin{eqnarray}
P_{n}^{m}(\xi)=(-1)^m\left(1-\xi^2 \right)^{\frac{m}{2}}\frac{1}{2^n n!}\frac{d^{n+m}}{d\xi^{n+m}}\left(\xi^2 -1 \right)^n.\,\,\,\,\,\,\,\label{leg}
\end{eqnarray}

\section{Two-dimensional geometry: the circle}
We suppose to consider a circular inhomogeneity of radius $R$ (conductivity $\sigma_2$) in the plane $(x,y)$ with conductivity $\sigma_{1}$. We consider an arbitrary applied (or pre-existing) potential $V_0(x,y)$ and we search for the perturbation induced by the inhomogeneity. Since the electric potential must be harmonic both inside and outside the interface, we have
\begin{eqnarray}
\label{pot2D}
V&=&V_0+\sum_{n=0}^{+\infty}\rho^n\left(A_n \cos n\vartheta+B_n \sin n \vartheta \right), \rho<R ,\\
\nonumber
 V&=&V_0+\sum_{n=0}^{+\infty}\rho^{-n}\left(\tilde{A}_n \cos n\vartheta+ \tilde{B}_n \sin n \vartheta \right), \rho>R ,
\end{eqnarray}
where $(\rho,\vartheta)$ are the standard polar coordinates. The potential $V_0$ and its derivatives $\frac{\partial V_0}{\partial \rho}$ can be expanded in Fourier series for $\rho=R$
\begin{eqnarray}
\label{svi2D}
V_0(R,\vartheta)=\sum_{n=0}^{+\infty}\left(C_n \cos n\vartheta+D_n \sin n \vartheta \right),\\
\nonumber
\frac{\partial V_0}{\partial \rho}(R,\vartheta)=\sum_{n=0}^{+\infty}\left(F_n \cos n\vartheta+ G_n \sin n \vartheta \right).
\end{eqnarray}
By substituting Eq.(\ref{pot2D}) in the anisotropic interface model (Eqs.(\ref{saltoV1}) and (\ref{saltoJ1})) and by using Eqs.(\ref{autocyl}) and (\ref{svi2D}), we obtain a set of equations for $A_n$ and $\tilde{A}_n$ 
\begin{eqnarray}
\nonumber
&&R^{-n}\tilde{A}_n-R^n A_n=r^+ \sigma_{1}\left(F_n-nR^{-n-1}\tilde{A}_n \right) \\
&&+r^- \sigma_{2}\left(F_n+nR^{n-1}{A}_n \right),
\\
\nonumber
&&\,\,\,\,\,\,\sigma_{2}\left(F_n+nR^{n-1}{A}_n \right)-\sigma_{1}\left(F_n-nR^{-n-1}\tilde{A}_n \right)\\
\nonumber
&&=gr^+\sigma_{1}R^{-2}n^2\left(F_n-nR^{-n-1}\tilde{A}_n \right)\\
&&\,\,\,\,\,\,-gR^{-2}n^2  \left(C_n+R^{-n}\tilde{A}_n \right),
\end{eqnarray}
and a similar one for the unknowns $B_n$ and $\tilde{B}_n$, not reported here for brevity. These systems can be easily solved obtaining the electrical potential in the whole plane. In the particular case of a uniform applied field $V_0=-xE_0=-\rho \cos \vartheta E_0$ only the coefficients $A_1$ and $\tilde{A}_1$ are different from zero and we obtain Eqs.(\ref{dentro}) and (\ref{fuori}) for $d=2$. A similar procedure (not reported here for brevity) can be followed to analyse the properties of the $\Pi$-model described by Eqs.(\ref{saltoV2}) and (\ref{saltoJ2}).

\section{Three-dimensional geometry: the sphere}
We consider now a spherical inhomogeneity of radius $R$ (conductivity $\sigma_2$) in a matrix with conductivity $\sigma_1$. As before, we assume an arbitrary applied potential $V_0(x,y,z)$ and we study the effects of the embedded particle. The final electric potential can be expanded as follows
\begin{eqnarray}
\label{pot3D}
V&=&V_0+\sum_{n=0}^{+\infty}\sum_{m=-n}^{+n}B_{nm}\rho^nY_{nm}(\vartheta,\varphi), \rho<R ,\\
\nonumber
 V&=&V_0+\sum_{n=0}^{+\infty}\sum_{m=-n}^{+n}C_{nm}\rho^{-n-1}Y_{nm}(\vartheta,\varphi), \rho>R, 
\end{eqnarray}
where we have introduced the spherical coordinates $(\rho,\vartheta,\varphi)$. The potential $V_0$ and its derivatives $\frac{\partial V_0}{\partial \rho}$ can be expanded in a series of spherical harmonics for $\rho=R$
\begin{eqnarray}
\label{svi3D}
V_0(R,\vartheta,\varphi)=\sum_{n=0}^{+\infty}\sum_{m=-n}^{+n}\beta_{nm}Y_{nm}(\vartheta,\varphi),\\
\frac{\partial V_0}{\partial \rho}(R,\vartheta,\varphi)=\sum_{n=0}^{+\infty}\sum_{m=-n}^{+n}\alpha_{nm}Y_{nm}(\vartheta,\varphi).
\nonumber
\end{eqnarray}
By substituting Eq.(\ref{pot3D}) in the anisotropic interface model (Eqs.(\ref{saltoV1}) and (\ref{saltoJ1})) and by using Eqs.(\ref{autosph}) and (\ref{svi3D}), we obtain a set of equations for $B_{nm}$ and $C_{nm}$ 
\begin{eqnarray}
\nonumber
&&\,\,\,\,\,\,R^{-n-1}C_{nm}-R^n B_{nm}\\
\nonumber
&&=r^+ \sigma_{1}\left[\alpha_{nm}-(n+1)R^{-n-2}C_{nm} \right] \\
&&\,\,\,\,\,\,+r^- \sigma_{2}\left[\alpha_{nm}+nR^{n-1}B_{nm} \right],
\\
\nonumber
&&\,\,\,\,\,\,\sigma_{2}\left[\alpha_{nm}+nR^{n-1}B_{nm} \right]\\
\nonumber
&&\,\,\,\,\,\,-\sigma_{1}\left[\alpha_{nm}-(n+1)R^{-n-2}C_{nm}\right]\\
\nonumber
&&=gr^+\sigma_{1}R^{-2}n(n+1)\left[\alpha_{nm}-(n+1)R^{-n-2}C_{nm} \right]\\
&&\,\,\,\,\,\,-gR^{-2}n(n+1)  \left[\beta_{nm}+R^{-n-1}C_{nm} \right].
\end{eqnarray}
It is now possible to find $B_{nm}$ and $C_{nm}$ obtaining the electrical potential in the whole space. In the particular case of a uniform applied field $V_0=-zE_0=-\rho \cos \vartheta E_0$ only the coefficients $B_{10}$ and $C_{10}$ are different from zero and we obtain Eqs.(\ref{dentro}) and (\ref{fuori}) for $d=3$.
We remark that a similar procedure can be followed for studying the dual interface described by Eqs.(\ref{saltoV2}) and (\ref{saltoJ2}).


\begin{thebibliography}{9}


\bibitem{torquato}
S. Torquato, \textit{Random Heterogeneous Materials} (Springer-Verlag, New
York, 2002).

\bibitem{milton}
G. W. Milton, \textit{The Theory of Composites} (Cambridge University Press,
Cambridge, 2002).

\bibitem{stratton}
J. A. Stratton,  {\it Electromagnetic theory} (Mc Graw Hill, New York, 1941).

\bibitem{landau}
L. D. Landau and E. M. Lifshitz, \textit{Electrodynamics of Continuous Media} (Pergamon Press, London, 1984).

\bibitem{benv-jmps}
Y. Benveniste, J. Mech. Phys. Solids \textbf{54}, 708 (2006).

\bibitem{benv-proc}
Y. Benveniste, Proc. R. Soc. London \textbf{A462}, 1593 (2006).

\bibitem{gu-he}
S.T. Gu, Q.C. He, J. Mech. Phys. Sol. \textbf{59}, 1413 (2011).

\bibitem{Kapitza}
P.L. Kapitza, \textit{Collected Papers of P.L. Kapitza}, 3 vol. (Pergamon Press, Oxford, 1964 - 1967).

\bibitem{benvmilo}
Y. Benveniste and T. Miloh, Int. J. Eng. Sci. \textbf{24}, 1537 (1986).

\bibitem{benv}
Y. Benveniste, J. Appl. Phys. \textbf{61}, 2840 (1987).

\bibitem{hasse}
D. P. H. Hasselman and L. F. Johnson, J. Compos. Mater. \textbf{21}, 508 (1987).

\bibitem{tor-rin}
S. Torquato and M. D. Rintoul, Phys. Rev. Lett. \textbf{75}, 4067 (1995).
 
\bibitem{lip1}
R. Lipton and B. Vernescu, Proc. R. Soc. London, Ser. A \textbf{452}, 329 (1996).

\bibitem{lip2}
R. Lipton and B. Vernescu, J. Appl. Phys. \textbf{79}, 8964 (1996).

\bibitem{chen}
H. Cheng and S. Torquato, Proc. R. Soc. London, Ser.  \textbf{A453}, 145 (1997).

\bibitem{nan}
C.-W. Nan, R. Birringer, D. R. Clarke, and H. Gleiter, J. Appl.Phys. \textbf{81}, 6692 (1997).

\bibitem{hashin}
 Z. Hashin, J. Appl. Phys. \textbf{89}, 2261 (2001).
 
\bibitem{duan-thermal}
H. L. Duan, B. L. Karihaloo, Phys. Rev. B \textbf{75}, 064206 (2007).
 
\bibitem{bonnet}
H. Le Quang, Q.-C. He, G. Bonnet, Phil. Mag. \textbf{91}, 3358 (2011).




\bibitem{lip3}
R. Lipton, SIAM J. Appl. Math. \textbf{57}, 347 (1997).

\bibitem{lip4}
R. Lipton, J. Mech. Phys. Solids \textbf{45}, 361 (1997).

\bibitem{moloh}
T. Miloh and Y. Benveniste, Proc. R. Soc. London, Ser. \textbf{A455}, 2687 (1999).

\bibitem{yvonnet}
J. Yvonnet, Q.-C. He, and C. Toulemonde, Compos. Sci. Technol. \textbf{68}, 2818 (2008).

\bibitem{bonnet-he}
H. Le Quang, G. Bonnet, Q.-C. He, Phys. Rev. B \textbf{81}, 064203 (2010).




\bibitem{eshelby1}
J. D. Eshelby,   Proc. R. Soc. London  \textbf{A241}, 376 (1957).

\bibitem{eshelby2}
J. D. Eshelby, Proc. R. Soc. London \textbf{A252}, 561 (1959).

\bibitem{walpole}
L. J. Walpole, Proc. R. Soc. London \textbf{A300}, 270 (1967).

\bibitem{taya}
H. Hatta and M. Taya, J. Appl. Phys. \textbf{58}, 2478 (1985).

\bibitem{lak}
W. S. Weiglhofer, A. Lakhtakia, and B. Michel, Microw. Opt. Technol. Lett. \textbf{15} 263 (1997).

\bibitem{giordano} 
S. Giordano and P. L. Palla, J. Phys. A: Math. Theor. \textbf{41}, 415205 (2008).

\bibitem{giorda}
S. Giordano, J. Electrost. \textbf{58}, 59 (2003).

\bibitem{giorda1}
S. Giordano and W. Rocchia, J. Appl. Phys. \textbf{98}, 104101 (2005).

\bibitem{appl1} 
B. D. Bertram, R. A. Gerhardt, and J. W. Schultz, J. Appl. Phys. \textbf{111}, 124913 (2012).

\bibitem{appl2}
T. Jeong, J.-G. Zhu, S. Chung, and M. R. Gibbons, J. Appl. Phys. \textbf{111}, 083510 (2012).

\bibitem{appl3}
D. Lee, S. D. Kang, H.-M. Kim, D.-H. Kang, H.-K. Lyeo, and K.-B. Kim, J. Appl. Phys. \textbf{111}, 073528 (2012), 

\bibitem{appl4}
S. H. Xie, Y. Y. Liu, and J. Y. Li, Appl. Phys. Lett. \textbf{92}, 243121 (2008).

\bibitem{appl5}
F. Gao, J. Qu, and M. Yao, J. Appl. Phys. \textbf{110}, 124314 (2011).

\bibitem{appl6}
H. B\"{o}hm and S. Nogales, Compos. Sci. Technol. \textbf{68}, 1181 (2008).

\bibitem{appl7}
S. Yu, S. Yang, and M. Cho, J. Appl. Phys. \textbf{110}, 124302 (2011).



\bibitem{maxwell}
J.C. Maxwell, \textit{A Treatise on Electricity and Magnetism} (Clarendon Press, Oxford, 1881).

\bibitem{prospe}
A. Prosperetti,\textit{Advanced Mathematics for Applications} (Cambridge University Press, Cambridge, 2011).

\bibitem{mori}
T. Mori and K. Tanaka, Acta Metallurgica \textbf{21}, 571 (1973).

\bibitem{benmiloneutrality}
Y. Benveniste and T. Miloh, J. Mech. Phys. Solids \textbf{47}, 1873 (1999).

\bibitem{michel}
B. Michel, A. Lakhtakia, W.S. Weiglhofer, T.G. Mackay,
Compos. Sci. Technol. \textbf{61}, 13 (2001).

\bibitem{epjb}
S. Giordano and P.L. Palla, Eur. Phys. J. B \textbf{85}, 59 (2012). 

\bibitem{lak1}
A. Lakhtakia, Microw. Opt. Technol. Lett. 17, 276 (1998).

\bibitem{gi1}
S. Giordano, Physica A \textbf{375}, 726 (2007).

\bibitem{gi2}
S. Giordano, J. Eng. Mater. Technol. \textbf{129}, 453 (2007).

\bibitem{lo}
R. M. Christensen and K. H. Lo, J. Mech. Phys. Solids \textbf{27}, 315 (1979).

\bibitem{mackay}
T. G. Mackay, J. Nanophotonics \textbf{5}, 051001 (2011).

\bibitem{jia1}
J. Qu, J. Appl. Mech. \textbf{60}, 1048 (1993).

\bibitem{jia2}
J. Qu, Mech. Mat. \textbf{14}, 269 (1993).

\bibitem{kar}
J. Wang, H. L. Duan, Z. P. Huang, and B. L. Karihaloo, Proc. R.
Soc. London, Ser. A \textbf{462}, 1355 (2006).

\bibitem{palla}
P. L. Palla, S. Giordano and L. Colombo, Phys. Rev. B \textbf{81}, 214113 (2010).

\bibitem{ropp}
L. Colombo and S. Giordano, Rep. Progr. Phys. \textbf{74}, 116501 (2011).



\bibitem{carmo}
M. P. do Carmo, \textit{Differential Geometry of Curves and Surfaces}
(Prentice-Hall, New York, 1976).

\bibitem{grad}
I. S. Gradshteyn and I. M. Ryzhik, \textit{Table of Integrals, Series and
Products} (Academic Press, San Diego, 1965).

\bibitem{abra}
M. Abramowitz and I. A. Stegun, \textit{Handbook of Mathematical
Functions} (Dover Publication, New York, 1970).





\end{thebibliography}
\end{document}